\documentclass[aps,prb,twocolumn,showpacs,groupedaddress]{revtex4}

\usepackage{graphicx}
\usepackage{dcolumn}

\begin{document}


\title{
Single-electron transistors in electromagnetic environments
}


\author{Michio Watanabe}
\homepage[]{http://www.riken.jp/lab-www/MQClab/michio/}
\altaffiliation{Present address: National Institute of Standards 
     and Technology, Division~817, 325 Broadway, Boulder, Colorado 80305; \\
               Email address: watanabe@boulder.nist.gov}
\affiliation{
Macroscopic Quantum Coherence Laboratory, Frontier Research System, 
RIKEN (The Institute of Physical and Chemical Research), 
2-1 Hirosawa, Wako, Saitama 351-0198, Japan
}


\date{September 23, 2003}

\begin{abstract}
The current--voltage ($I$--$V$) characteristics of 
single-electron transistors (SETs) have been measured 
in various electromagnetic environments.  
Some SETs were biased with one-dimensional arrays 
of dc superconducting quantum interference 
devices (SQUIDs).  
The purpose was to provide the SETs 
with a magnetic-field-tunable environment 
in the superconducting state, and a high-impedance 
environment in the normal state.  
The comparison of SETs with SQUID arrays 
and those without arrays in the normal state 
confirmed that the effective charging energy of SETs 
in the normal state becomes larger in the 
high-impedance environment, as expected 
theoretically.  
In SETs with SQUID arrays in the superconducting state, 
as the zero-bias resistance of the SQUID arrays 
was increased to be much larger than the quantum 
resistance $R_K\equiv h/e^2\approx26$~k$\Omega$, 
a sharp Coulomb blockade was induced, and the current 
modulation by the gate-induced charge was changed 
from $e$ periodic to $2e$ periodic 
at a bias point $0<|V|<2\Delta_0/e$, where $\Delta_0$ 
is the superconducting energy gap.  
The author discusses the Coulomb blockade 
and its dependence on the gate-induced charge 
in terms of the single Josephson junction with 
gate-tunable junction capacitance.  
\end{abstract}

\pacs{73.23.Hk, 74.50.+r, 74.78.Na\\
Phys. Rev. B {\bfseries 69}, 094509 (2004).\\
DOI: 10.1103/PhysRevB.69.094509}

\maketitle

\section{Introduction}
\label{sec:intro}
The single-electron transistor (SET), which 
consists of two small-capacitance tunnel junctions 
in series and a gate electrode, is an important example 
of the single-electron-tunneling circuit.  
Although the SET is more complex than the single junction, 
the SET has an advantage over the single junction 
in that the single-electron charging 
effects typified by ``Coulomb blockade" are easily 
observed in the SET.  
As a result,    
the SET can be used, e.g., as an electrometer with 
an extremely high sensitivity, $\ll e/\sqrt{\mbox{Hz}}$   
(e.g., Ref.~\onlinecite{Ave91}).    
When the electrodes of the tunnel junctions in the SET 
are in the superconducting state, the SET is also 
important from the viewpoint of quantum computing;  
the first one-quantum-bit operation in 
solid-state electronic devices was demonstrated 
in a circuit with small-capacitance Josephson 
tunnel junctions, and the state of the quantum bit 
was read out by measuring the Josephson-quasiparticle 
(JQP) current of a superconducting SET.\cite{Nak99}    
The supercurrent of superconducting SETs 
has also been studied extensively.\cite{Joy95} 
One expects that the supercurrent is $2e$ periodic 
in the gate-induced charge.  
In many experiments, however, 
a period of $e$ has been seen, which suggests the existence 
of subgap quasiparticle states,\cite{Joy95} 
or ``quasiparticle poisoning."   
The periodicity of the current was also discussed 
at finite voltages.\cite{Gee90PRL,Tuo92,Ama94}  
The periodicity depends strongly on how the leads 
connected to the sample are filtered,\cite{Joy95} so 
that the ``quasiparticle poisoning" would be 
viewed as an environmental effect.  

Theoretically,\cite{Ing92} 
the details of single-electron charging 
effects in SETs depend on the impedance  
of the electromagnetic environment, $Z(\omega)$,  
both in the normal state and in the superconducting state.  
It is predicted that as $\mbox{Re}[Z(\omega)]$ 
is increased, the voltage scale of the Coulomb 
blockade in the normal-state current--voltage 
($I$--$V$) characteristics becomes larger, 
and in the superconducting state, the supercurrent 
is replaced by the Coulomb blockade.  
The experiments\cite{Joy95,Gee90PRL,Tuo92,Ama94} 
on superconducting SETs mentioned above were done in 
the low-impedance environment, where   
$\mbox{Re}[Z(\omega)]$ was much smaller 
than the quantum resistances: 
$R_Q\equiv h/(2e)^2\approx6.5$~k$\Omega$ 
for Cooper pairs and 
$R_K\equiv h/e^2\approx26$~k$\Omega$ 
for quasiparticles.  
In order to obtain higher $\mbox{Re}[Z(\omega)]$, 
thin-film resistors\cite{Hav91,Cle92} can be used 
for the on-chip leads.  
A superconducting SET was biased with 
$\approx$50~k$\Omega$ Cr resistors,   
and a Coulomb blockade was observed 
in the $I$--$V$ curve.\cite{Hav94}  
Similar resistors 
($2-20$~k$\Omega$)  
were also employed for the study of Cooper-pair 
cotunneling in superconducting SETs.\cite{Lot03}  
A drawback of these thin-film resistors is that 
the resistance is not tunable, i.e., 
the same SET cannot be measured in different 
environments.   
A tunable environment has been realized 
by capacitively coupling a two-dimensional 
electron gas to a SET.\cite{Kyc01,Rim02}  
In the work presented here, a simpler method is employed  
to create tunable environments for superconducting SETs.  

We use one-dimensional (1D) arrays of dc 
superconducting quantum interference devices 
(SQUIDs) for the on-chip leads,\cite{Wat01PRL,Wat03}  
which can be fabricated simultaneously with 
superconducting SETs.  
The SQUID configuration enables us to 
vary {\it in situ} the effective impedance of 
the arrays by applying a weak external magnetic field 
($1-10$~mT) perpendicular to the SQUID loop.  
The superconducting SET in our samples, on the 
other hand, does not have a SQUID configuration, 
and therefore its parameters are practically 
independent of the external magnetic field.  
The zero-bias resistance 
of SQUID arrays at low temperatures can 
be controlled over several orders of magnitude 
(e.g., Fig.~13 of Ref.~\onlinecite{Wat03}), and    
SQUID arrays are especially suitable 
for achieving a high-impedance environment.  
In fact, distinct 
Coulomb blockade was observed in single Josephson 
junctions by biasing with SQUID arrays.\cite{Wat01PRL}   
When the sample is driven to the normal state, 
the array leads are no longer tunable, but the 
resistance can be much larger than $R_K$.  
Hence, it is still possible to study the 
environmental effects on SETs by comparing 
the samples with and without array leads.  
In this paper, the environmental effects on SETs 
is discussed both in the normal state and in the superconducting state.  
An emphasis is placed on the case of a high-impedance environment, 
which has not been thoroughly experimentally studied.

\section{Experiment}
\subsection{Fabrication of small-capacitance tunnel junctions}
The tunnel junctions (Al/Al$_2$O$_3$/Al) were fabricated 
on a SiO$_2$/Si substrate with Au/Ni bonding pads.   
We employed a process based on electron-beam lithography 
and double-angle shadow evaporation, which is similar 
to the one described in Ref.~\onlinecite{Wat03}.  
The evaporation of Al was done at an rate of 
$0.1-0.2$~nm/s in a vacuum system 
with the base pressure of $\leq10^{-8}$~Pa.  
During the evaporation, the pressure was 
usually $(2-4)\times10^{-6}$~Pa.  
The Al$_2$O$_3$ tunnel barrier was formed by exposing 
the base Al layer to $1-20$~Pa of O$_2$ for $0.5-2$~min.\ 
before the deposition of the top Al layer.   
The thickness of the barrier determines 
the normal-state resistance of the junction per unit 
junction area.   The normal-state resistance  
is a key parameter of the sample, and   
will be discussed in Sec.~\ref{sec:charac}.

\begin{figure}
\includegraphics[width=0.95\columnwidth,clip]{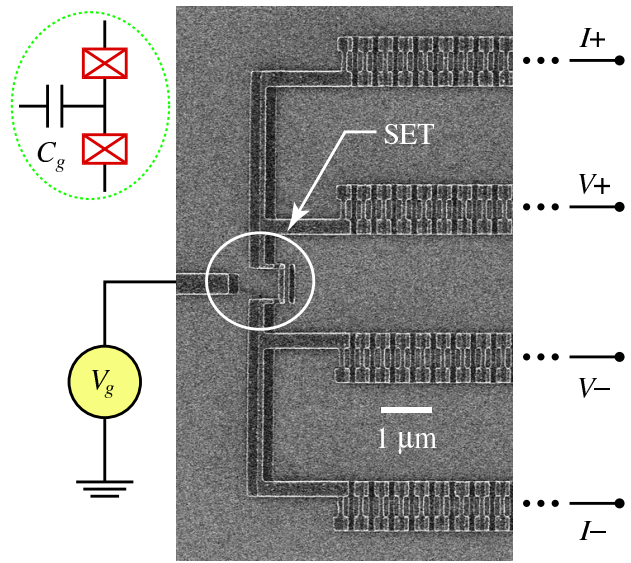}
\caption{\label{fig:SEM}
(Color online)
A scanning electron micrograph 
of an Al/Al$_2$O$_3$/Al single-electron transistor (SET) 
with SQUID-array leads.
An equivalent circuit for the SET is shown in the top left 
corner.  
}
\end{figure}
Figure~\ref{fig:SEM} shows a scanning electron micrograph 
of a sample. 
The SET is biased with two pairs of leads 
enabling four-point measurements.   
A part of each lead close to the SET consists of a 1D array of 
dc SQUIDs.  

\subsection{Measurement at low temperatures}
The samples were measured 
in a $^{3}$He--$^{4}$He dilution refrigerator 
(Oxford Instruments, Kelvinox 400) mainly at $T=0.02-0.6$~K, 
and the normal-state resistance was determined at $T=1.8-4.5$~K.   
(Note that the superconducting transition temperature of Al is 1.2~K.)  
The temperature was calculated from the resistance 
of a ruthenium-oxide thermometer fixed to the mixing chamber.  
Magnetic fields on the order of $1-10$~mT were applied by means 
of a superconducting solenoid.  In this magnetic-field range, 
the temperature error of ruthenium-oxide thermometers due to 
magnetoresistance is negligibly small (e.g., a typical value 
of the error at $T=0.05$~K is less than 0.1\%).\cite{Wat01Cryo}
The samples were placed inside a copper can which was
thermally connected to the mixing chamber.  
Because there was no low-temperature filtering, 
we inserted low-pass filters between the cables 
connected to the cryostat and the measurement circuit.  

The $I$--$V$ curve of the SET was measured 
in a four-point configuration (see Fig.~\ref{fig:SEM}).   
The bias was applied through one-pair of leads,  
and the potential difference was measured through the other pair 
of leads with a differential voltage amplifier  
(DL Instruments, 1201, $>1$~G$\Omega$ input impedance).  
The current was measured with a current amplifier 
(DL Instruments, 1211).  
The SQUID arrays could be measured in a two-point configuration 
(same current and voltage leads). 
When we measure the arrays on the same side of the SET 
(e.g., $I+$ and $V+$ in Fig.~\ref{fig:SEM}), 
current does not flow through the SET, 
and the series resistance of two arrays is obtained.  
The zero-bias resistance measured in this configuration 
will be discussed in Sec.~\ref{sec:charac}.

\subsection{Characterization of the samples}  
\label{sec:charac}
In this work, we measured the three pairs of SETs 
listed in Table~\ref{tab:sample}.
\begin{table}
\caption{
List of the samples.  
$R_n$ is the normal-state resistance  
of the single-electron transistor,  
$r_n'$ is the normal-state resistance per junction 
pair of the SQUID-array leads, 
and $C_g$ is the capacitance between the island electrode 
and the gate electrode.     
Samples~1b, 2b, and 3b do not have SQUID arrays in the leads.  
}
\label{tab:sample}
\begin{ruledtabular}
\begin{tabular}{cdcc}
Sample & \multicolumn{1}{c}{$R_n$ (k$\Omega$)} & 
$r_n'$ (k$\Omega$) & $C_g$ (aF)\\ 
\hline
1a &  82 & $-$ & 6.0 \\
1b & 102 & 5.6 & 6.0 \\
2a &  57 & $-$ & 4.8 \\
2b &  85 & 4.1 & 4.9 \\ 
3a &  17 & $-$ & 4.8 \\ 
3b &  23 & 1.4 & 4.8 \\ 
\end{tabular}
\end{ruledtabular}
\end{table} 
Each pair (e.g., samples~1a and 1b) was fabricated 
simultaneously on the same chip.  
One SET has SQUID-array leads with $N=65$ junction pairs 
in each array, while the other SET has no SQUID-array 
leads ($N=0$).  
For all the samples, the junction area was designed 
to be 0.1$\times$0.1~$\mu$m$^2$ in the SET 
and 0.3$\times$0.1~$\mu$m$^2$ in the SQUID arrays.    

The uniformity of the SQUID arrays could be estimated 
by measuring the normal-state resistance 
in all two-point configurations (six in total).    
In some configurations, current flows through the SET
 and thus the normal-state resistance $R_n$ of the SET 
has to be subtracted from the measured resistance in order to obtain 
the series resistance of two arrays.  
For our samples, the results of the two-point measurements 
agreed with one another within 3\%--4\%.  
By averaging all the results, the normal-state resistance 
$r_n'$ per junction pair of the arrays is calculated, 
and shown in Table~\ref{tab:sample}.  
If the tunnel-barrier thickness is identical 
for all junctions on the same chip, one would 
expect $R_n/r_n'\approx12$ from the junction area.   
From Table~\ref{tab:sample}, the ratio is $16-21$, 
which is the correct order of magnitude.

\begin{figure}
\includegraphics[width=0.95\columnwidth,clip]{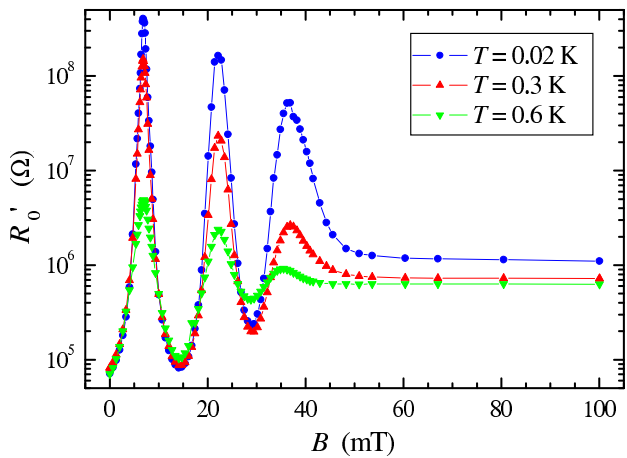}
\caption{\label{fig:R0leads}
(Color online) 
Zero-bias resistance of two SQUID-array leads connected 
in series on the same side 
of the SET vs external magnetic field for sample~2b 
at $T=0.02$, 0.3, and 0.6~K. 
}
\end{figure}

From the normal-state resistance, I calculate 
the Josephson energy, 
which is an important parameter of the samples.  
For example, the maximum Josephson energy $E_{J0}'$ 
between adjacent islands 
in the SQUID arrays is given by the Ambegaokar-Baratoff formula, 
\begin{equation}
E_{J0}'=\frac{\,h\Delta_0\,}{\,8e^2r_n'\,}~,   
\end{equation}
where $\Delta_0$ is the superconducting energy gap 
($\approx 0.2$~meV for Al).  
Because of the SQUID configuration, the effective Josephson energy 
$E_J'$ in the arrays is modulated periodically by applying an external 
magnetic field $B$ perpendicular to the SQUID loop,  
\begin{equation}
\label{eq:EJ'}
E_J'=E_{J0}'\left|\cos\left(\pi
\frac{\,BA\,}{\,\Phi_0\,}\right)\right|~,  
\end{equation}
so long as $B$ is sufficiently smaller than the critical field,    
where $\Phi_0= h/2e= 2\times10^{-15}$~Wb 
is the superconducting flux quantum, and $A$ is the effective area 
of the SQUID loop, which is 0.7$\times$0.2~$\mu$m$^2$ in our samples.   
Equation~(\ref{eq:EJ'}) is a key to understand 
Fig.~\ref{fig:R0leads}, which shows that the zero-bias 
resistance $R_0'$ of two SQUID arrays in series oscillates 
as a function of $B$ until $B$ becomes 
comparable to the critical field.  
Note that the first peak appears at $B=7$~mT, 
where the normalized flux $BA/\Phi_0$ is 0.5 
for our samples with $A=0.14$~$\mu$m$^2$. 
In Ref.~\onlinecite{Hav00}, the SQUID array was modeled as 
a network of capacitors and inductors, and the real part of 
the array impedance was shown to be proportional to 
$1/\sqrt{E_J'}$ at low enough frequencies.  
At $B\geq60$~mT in Fig.~\ref{fig:R0leads}, 
the superconductivity is suppressed, 
and $R_0'$ takes a value comparable to $2Nr_n'$.  
The small temperature dependence in this magnetic-field 
regime is due to the single-electron charging effect.  
In general, the $I$--$V$ curve of SQUID arrays 
is nonlinear (e.g., Fig.~3 of Ref.~\onlinecite{Wat03}), 
and therefore, SQUID arrays are not described by a linear 
impedance model.\cite{Hav00}  
However, it would be a good approximation\cite{Wat03} 
to use $R_0'$ for characterizing the electromagnetic 
environment of the SET.  
I also assume that the SET is in a high-impedance 
environment when $R_0'\gg R_K$.     
As for the samples without SQUID arrays, the resistance of 
the leads is always much smaller than $R_K$, and the SET 
is in a low-impedance environment.    
The Josephson energy in the SET is calculated 
in a similar way, but note that it does not 
oscillate as a function of $B$ because the junctions 
in the SET do not have a SQUID configuration.  

Another important parameter is the charging energy, 
which is inversely proportional to the capacitance of 
the junction.  
The capacitance can be estimated from the junction area 
with a specific capacitance on the order of $10^2$~fF/$\mu$m$^2$.  
In our samples, the junction area is much larger, 
or the charging effect is much weaker, in the SQUID arrays 
than in the SET.  
In the normal state, for example, the $I$--$V$ curve of the SQUID 
arrays is almost linear (data not shown, similar to 
the bottom curve in Fig.~3 of Ref.~\onlinecite{Wat03}) 
even when the SET shows a well developed Coulomb blockade 
like in Fig.~\ref{fig:IVnormal}(b),    
which would be favorable when the arrays are used as leads.

\begin{figure}
\includegraphics[width=0.9\columnwidth,clip]{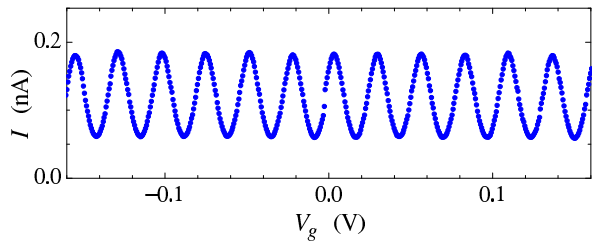}
\caption{\label{fig:IGnormal}
(Color online)
Modulation of the current by the gate voltage 
in the normal state ($B=0.1$~T) at $T=0.02$~K for sample~1a.   
The bias voltage is 0.04~mV.  
}
\end{figure}

The capacitance $C_g$ between the island electrode 
and the gate electrode is determined by the period 
of the gate modulation in the normal state.  
An example is shown in Fig.~\ref{fig:IGnormal}.   
Though the design of the SET is the same for all the samples, 
the values of $C_g$ in Table~\ref{tab:sample}  
are divided into two groups: 6~aF for samples~1a 
and 1b, and $<$5~aF for the others.  
The reason could be that we used two SiO$_2$/Si wafers  
whose thickness $t$ of the SiO$_2$ layer is different, 
and that we fabricated samples~1a and 1b on one wafer 
($t=0.2$~$\mu$m) and samples~2a--3b on the other ($t=0.5$~$\mu$m).  

\section{Results and discussion}
\label{sec:RD}
\subsection{Single-electron transistors in the normal state}
\label{subsec:normal}
\begin{figure}
\includegraphics[width=0.6\columnwidth,clip]{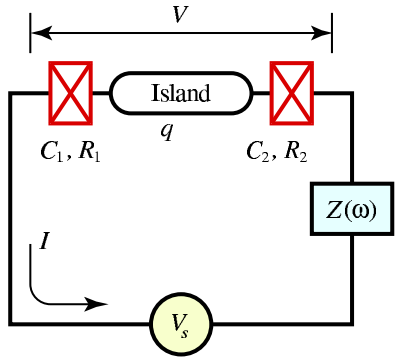}
\caption{\label{fig:diagram}
(Color online)
Double-junction system connected to a voltage source $V_s$ 
via electromagnetic environment $Z(\omega)$.  
}
\end{figure}

At $T=0$, the $I$--$V$ curve of the SET in the normal state 
is expected to have a region where $I=0$.  
Such a region is called a ``Coulomb gap," and depends not only 
on the gate voltage $V_g$ 
but also on the electromagnetic environment.\cite{Ing92}  
The circuit considered in the theory\cite{Ing92} is shown 
in Fig.~\ref{fig:diagram}, 
where $C_i$ and $R_i$ ($i=1,2$) 
are the capacitance and the tunnel resistance, respectively, 
for the $i$th junction, 
and $Z(\omega)$ is the impedance of 
the electromagnetic environment.   
The theory assumes that $R_i\gg R_K$ and $R_i \gg \mbox{Re}[Z(\omega)]$.    
In Fig.~\ref{fig:diagram}, the gate is not drawn explicitly 
because when $C_i\gg C_g$, which is the case in our samples, 
the influence of the gate can be 
included in the effective island charge, 
\begin{equation}  
\label{eq:q}
q=ne+C_gV_g+Q_0, 
\end{equation}
where $n$ is an integer 
and $Q_0$ is the background charge.

In the theory, the drift of $Q_0$ is not considered, and thus 
for simplicity, $Q_0=0$ in Eq.~(\ref{eq:q}).  
In the experiment, however, the drift often becomes a problem.     
In samples~1a, 1b, 3a, and 3b, the drift 
was usually $<0.1e/\mbox{day}$, which is negligible   
for the discussion in this paper.  
In samples~2a and 2b, on the other hand, 
the drift was not slow enough 
($\sim e/\mbox{hour}$) compared to the time needed 
to measure hundreds of $I$--$V$ curves at different values 
of $V_g$ in our experimental setup.  
Thus, for samples~2a and 2b, we mainly measured $I$ vs $V_g$ 
keeping the source voltage $V_s$ constant.

\begin{figure}
\includegraphics[width=0.95\columnwidth,clip]{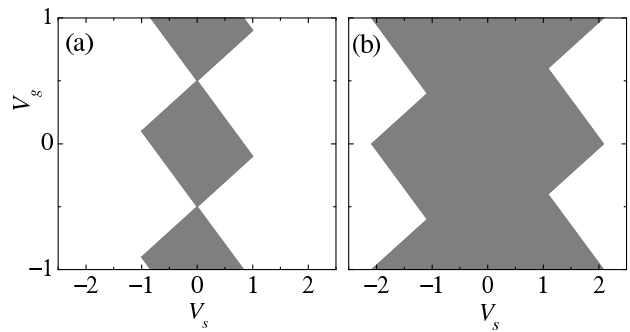}
\caption{\label{fig:diamond}
The theoretical region of zero current at $T=0$ 
(Coulomb gap) for single-electron transistors  
with $C_2/C_1=1.5$  
in (a) the low-impedance environment 
and in (b) the high-impedance environment.  
The source voltage $V_s$ is in units of $e/(C_1+C_2)$ and    
the gate voltage $V_g$ is in units of $e/C_g$.   
}
\end{figure}

\begin{figure}
\includegraphics[width=0.95\columnwidth,clip]{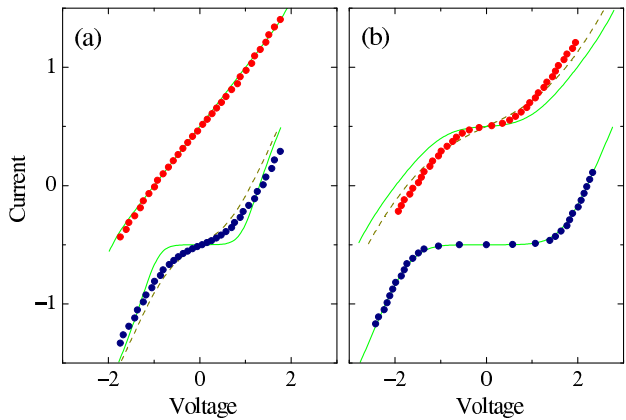}
\caption{\label{fig:IVnormal}
(Color online)
Current--voltage characteristics of single-electron 
transistors in the normal state  
in (a) low-impedance environment for sample~1a 
and in (b) high-impedance environment for sample~1b.  
The current and the voltage are in units of 
$e/(R_1+R_2)(C_1+C_2)$ and $e/(C_1+C_2)$, respectively.  
The solid circles are experimental data at $B=0.1$~T 
and $T=0.02$~K, 
where $R_1+R_2=R_n$ and $C_1+C_2=1$~fF are employed.    
The solid (dashed) curves are the theoretical prediction 
for $k_BT/E_C=0.1$ (0.3), $R_1/R_2=1.0$, and $C_1/C_2=1.0$.  
The upper (lower) data sets are for $q=0.5$ (0.0).  
The origin of the current axis 
is offset for each curve for clarity.    
}
\end{figure}

Figure~\ref{fig:diamond} shows the theoretical Coulomb gap 
on the $V_g$--$V_s$ plane for $Z(\omega)=0$ (low-impedance environment) 
and for $\mbox{Re}[Z(\omega)]\gg R_K$ (high-impedance environment).  
The Coulomb gap is periodic in $V_g$ for both the cases.  
However, the magnitude of the gap is different, and e.g.,   
at $V_g=0$, $\pm e/C_g,$ $\pm 2e/C_g,\ldots,$ the gap 
in the high-impedance environment is about twice as large 
as that in the low-impedance environment.  
It is important to note a qualitative difference 
at $V_g=\pm0.5e/C_g,$ $\pm1.5e/C_g,$ $\pm2.5e/C_g,\ldots$~. 
The gap vanishes in the low-impedance environment, 
but survives in the high-impedance environment.  
These theoretical predictions have been confirmed in 
our experiments.  
Figure~\ref{fig:IVnormal} shows the normal-state 
$I$--$V$ curves at $T=0.02$~K for samples~1a and 1b.  
We drove the samples into the normal state 
by applying a magnetic field of 0.1~T 
perpendicular to the substrate.  
In the normal state, the SQUID arrays of sample~1b 
have $R_0'=1.4$~M$\Omega$ $\gg R_K$ at $T=0.02$~K, 
and thus, I assume that the SET in sample~1b is 
in a high-impedance environment.  
The SET in sample~1a, on the other hand, is 
in a low-impedance environment because it is not 
biased with the arrays.  
In the lower data set ($q=0.0$), the size of the Coulomb gap in 
Fig.~\ref{fig:IVnormal}(b) is about twice as large 
as that in Fig.~\ref{fig:IVnormal}(a). 
Moreover, in the upper data set ($q=0.5$), the $I$--$V$ curve 
in Fig.~\ref{fig:IVnormal}(a) 
is linear, i.e., no Coulomb gap, while the curve 
in Fig.~\ref{fig:IVnormal}(b)      
still shows a considerable nonlinearity.

I have also calculated the $I$--$V$ curves 
based on the theory 
assuming $R_1+R_2=R_n$.    
For the lower data set in Fig.~\ref{fig:IVnormal}(b), 
a perfect agreement is obtained 
with $R_1/R_2=1.0$, $C_1/C_2=1.0$, $C_1+C_2=1$~fF, 
and $k_BT/E_C=0.1$ ($T=0.09$~K), 
where $E_C=e^2/2(C_1+C_2)$.  
These parameters are employed for all the solid 
curves in Fig.~\ref{fig:IVnormal}.    
The solid curve does not always reproduce 
the experimental data very well, but I 
nevertheless conclude that our experiments 
have demonstrated the environmental effects 
predicted by the theory. 
I emphasize that the solid curves are calculated 
with the same set of parameters, and 
the discrepancy is reduced if 
the parameters are adjusted for the calculation 
for each data set.  
The dashed curves for the lower data set 
of Fig.~\ref{fig:IVnormal}(a) 
and for the upper data set of Fig.~\ref{fig:IVnormal}(b), 
which agree better with the experimental data,  
are obtained by raising the temperature in the calculation 
to $k_BT/E_C=0.3$ ($T=0.28$~K), keeping the other 
parameters constant.

\begin{figure}
\includegraphics[width=0.95\columnwidth,clip]{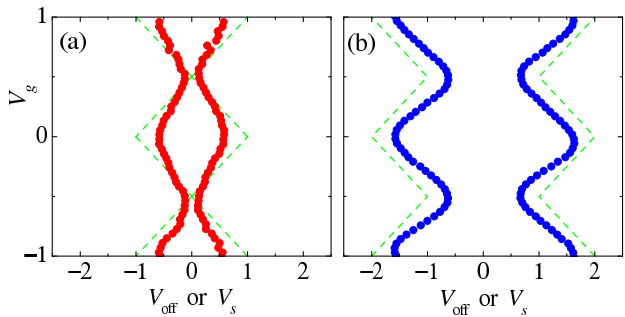}
\caption{\label{fig:Voff}
(Color online)
Offset voltage defined in Eq.~(\ref{eq:Voff}) 
as an estimate of the Coulomb gap for   
(a) sample~1a and (b) sample~1b.  
The broken lines are the theoretical predictions  
for (a) low-impedance environment and 
(b) high-impedance environment.  
The units of the axes are the same as in Fig.~\ref{fig:diamond}, 
and the parameters, $R_1$, $R_2$, $C_1$, and $C_2$, are 
the same as in Fig.~\ref{fig:IVnormal}.   
}
\end{figure}

At finite temperatures, nonzero current flows even within 
the Coulomb gap.  Therefore, it is not straightforward 
to determine the Coulomb gap from the experiment.  
As an estimate of the Coulomb gap, 
I have calculated the offset voltage $V_{\rm off}$ 
by fitting 
\begin{equation}
\label{eq:Voff}
I \propto (V-V_{\rm off}) 
\end{equation}
to the data in the high-bias regime, 
$I\geq 0.3e/(R_1+R_2)(C_1+C_2)$ for $V_{\rm off}>0$ and 
$I\leq -0.3e/(R_1+R_2)(C_1+C_2)$ for $V_{\rm off}<0$.   
The results for samples~1a and 1b are plotted in Fig.~\ref{fig:Voff} 
together with the theoretical predictions for the 
low-impedance environment and high-impedance environment, 
respectively.  
Here, I used the same parameters as in Fig.~\ref{fig:IVnormal}.  
The experimental $V_{\rm off}$ is consistent with the theoretical 
prediction.  

In fact, one of the assumptions in the theory 
for the high-impedance case, $R_i\gg \mbox{Re} 
[Z(\omega)]$, is not fulfilled in the experiment.  
In the normal state, $R_n<R_0'$ in sample~1b, 
so that the SET in sample~1b is current-biased.     
However, I am still convinced that the 
comparison with the theory is meaningful,  
because we measured the SET in four-terminal 
configuration, and because the measured 
$I$--$V$ curves are qualitatively explained 
within the theory.  
Furthermore, the parameters used in the calculations 
for Fig.~\ref{fig:IVnormal} are reasonable.   
It is common to observe that the effective electron temperature 
becomes considerably higher than 
the temperature of the mixing chamber 
in the experiments of small-capacitance tunnel 
junctions.\cite{Wat03}  
The temperature difference is likely to be large 
when the cryostat leads are not filtered at low temperatures, 
which is the case in our cryostat.  
Most importantly, $C_1+C_2=1$~fF is consistent 
with our junction size, and as I mentioned earlier, 
much larger than $C_g$.  
This value of the total capacitance will be 
used again in the discussion in Sec.~\ref{subsec:super}.   

\subsection{Single-electron transistors in the superconducting state}
\label{subsec:super}

\begin{figure}
\includegraphics[width=0.9\columnwidth,clip]{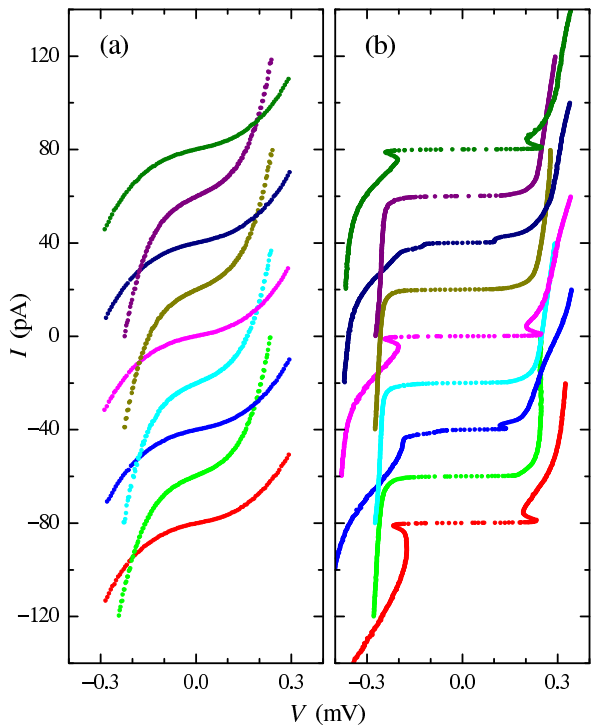}
\caption{\label{fig:IVsuper}
(Color online) 
Sets of the current--voltage curves 
of the same superconducting SET (sample~1b)  
in different environments: 
(a) $R_0'=0.2$~M$\Omega$ ($B=0$), 
and (b) $R_0'=0.3$~G$\Omega$ ($B=6.8$~mT), 
where $R_0'$ is the zero-bias resistance 
of two SQUID-array leads connected in series.      
Both in (a) and (b), from bottom to top, 
the normalized gate-induced charge $(C_gV_g+Q_0)/e$ 
increases from $-2.0$ to $+2.0$ in steps of 0.5.  
The origin of the current axis is offset 
for each curve for clarity.  
All the curves were measured at $T=0.02$~K.   
}
\end{figure}

For the samples with SQUID-array leads,   
the same superconducting SET can be studied 
in different electromagnetic 
environments by changing the external magnetic field 
on the order of a few mT.  
In this work, I look at the low-bias region of 
$|V|<2\Delta_0$, where the $I$--$V$ curve is sensitive to 
the state of the electromagnetic environment.   
In all the samples with SQUID-array leads,  
the $I$--$V$ curve of the superconducting SET 
developed a Coulomb blockade as $R_0'$ was increased 
by tuning the field.    
Figure~\ref{fig:IVsuper} shows the $I$--$V$ curves of 
the SET in sample~1b in two different environments.  
As I mentioned earlier, 
the parameters of the SET should be independent 
of the field, because the SET does not 
have a SQUID configuration and the field applied 
in Fig.~\ref{fig:IVsuper} is much smaller 
than the critical field. 
The electromagnetic environment for the SET  
(the SQUID arrays), on the other hand, is strongly 
varied with the field as we have seen in 
Fig.~\ref{fig:R0leads}.  
The behavior of the SET demonstrated 
in Fig.~\ref{fig:IVsuper} does not result from the 
magnetic-field influence on the $I$--$V$ curve of the 
SET, but rather from an environmental effect on the SET.  
I note here that the SETs without SQUID arrays were 
also measured at $B=0$ and at $B\approx7$~mT, 
and that the $I$--$V$ curves were almost the same.  
The well-developed Coulomb blockade 
in Fig.~\ref{fig:IVsuper}(b) indicates that 
the SET is in a high-impedance environment, 
which is consistent with $R_0'=0.3$~G$\Omega$ $\gg R_Q$. 
From the viewpoint of $R_0'$,  
the SET is not in a low-impedance environment 
in Fig.~\ref{fig:IVsuper}(a) because 
$R_0'=0.2$~M$\Omega$ is already larger than $R_Q$.  
Thus, in order to study the low-impedance case, 
we have to measure samples without SQUID-array leads.     
Rather surprisingly, e.g., in sample~1a, the $I$--$V$ 
curve and its dependence on the gate-induced charge 
were qualitatively almost the same 
as Fig.~\ref{fig:IVsuper}(a), and the supercurrent 
was too small to be seen in the $I$--$V$ curve.  
This is due to ``quasiparticle poisoning," 
which is likely to occur in our cryostat which has 
no low-temperature noise filters.  
In addition, our biasing scheme is not ideal for 
the supercurrent measurement because  
the samples without SQUID-array are voltage biased.  
The supercurrent was not detected in samples~2a or 3a, 
either.

\begin{figure}
\includegraphics[width=0.95\columnwidth,clip]{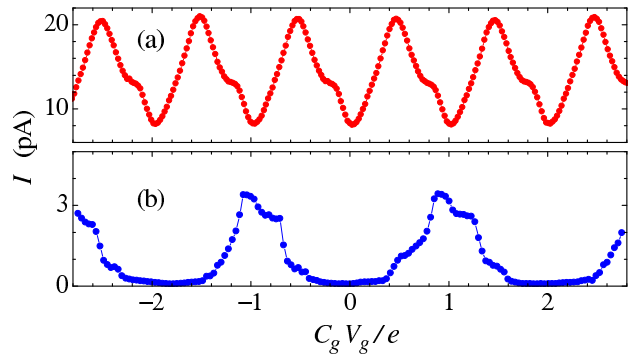}
\caption{\label{fig:IGsuper}
(Color online) 
Modulation of the current by the gate-induced charge  
in the same SET as in Fig.~\ref{fig:IVsuper}.   
The potential drop across the SET is 0.15~mV both in 
(a) and (b).  The state of the environment for the SET 
in (a) and (b) is the same as in Fig.~\ref{fig:IVsuper}(a) 
and Fig.~\ref{fig:IVsuper}(b), respectively.    
}
\end{figure}

``Quasiparticle poisoning" is also suggested 
by the $e$ periodicity in the current modulation 
by the gate-induced charge, which was observed 
in all samples without SQUID-array leads.  
The $e$ periodicity was also found in sample~1b 
when $R_0'=0.2$~M$\Omega$, as shown 
in Fig.~\ref{fig:IGsuper}(a), where the potential 
drop across the SET is 0.15~mV.  
When $R_0'$ is increased to 0.3~G$\Omega$ 
[Fig.~\ref{fig:IGsuper}(b)], it was replaced 
by $2e$ periodicity, which indicates  
Cooper-pair transport.    
The SQUID arrays with 
sufficiently large $R_0'$ acted as a filter 
that suppresses the contribution of 
quasiparticles to the charge transport.  
This change in the periodicity was also 
found in sample~2b.  In sample~3b, however, 
only $e$ periodicity was seen probably 
because the maximum of $R_0'$ was only 9~M$\Omega$.  
The curve in Fig.~\ref{fig:IGsuper}(b) is 
calculated from the $I$--$V$ curves, 
because we could not fix the 
potential drop across the SET when it is much 
smaller than that across the SQUID arrays 
in our experimental setup.  

\begin{figure}
\includegraphics[width=0.95\columnwidth,clip]{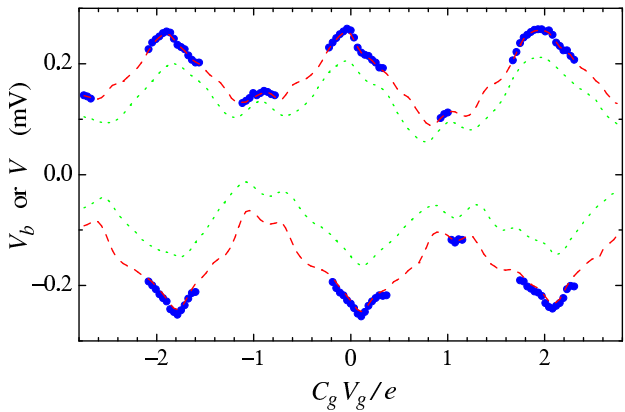}
\caption{\label{fig:Vb}
(Color online) 
Blockade voltage $V_b$ as a function 
of the gate-induced charge in the same SET as 
in Figs.~\ref{fig:IVsuper} and \ref{fig:IGsuper}.     
The state of the environment is the same as in 
Figs.~\ref{fig:IVsuper}(b) and \ref{fig:IGsuper}(b).  
The dotted (broken) curves denote the 
potential drop across the SET at $I=\pm0.2$~pA  
($\pm0.8$~pA). } 
\end{figure}

Below I focus on the case of $R_0'\gg R_Q$,  
and discuss the Coulomb blockade 
of Cooper-pair tunneling.  
In the high-impedance environment, 
a current-biased single Josephson junction 
is expected theoretically\cite{Ave91,Sch90} 
to have a ``back-bending" $I$--$V$ curve, 
which has been experimentally\cite{Hav91,Wat01PRL,Wat03} 
confirmed.  The ``back-bending" is also seen 
in Fig.~\ref{fig:IVsuper}(b) for certain values of 
the gate-induced charge.  
When the ``back-bending" is clearly seen 
in the SET $I$--$V$ curve, 
I define the blockade voltage $V_b$ as 
the local voltage maximum (minimum) for $V_b>0$ 
($V_b<0$) in the low-current part of the $I$--$V$ 
curve.  The measured $V_b$ is plotted in Fig.~\ref{fig:Vb} 
as a function of the gate-induced charge.  
The dotted and broken curves represent $V$ at 
constant $I$ calculated from the $I$--$V$ curves.  
These curves would also characterize the Coulomb 
blockade.  
The analytic expression of $V_b$ at low 
temperatures has been obtained 
for single Josephson junctions as  
\begin{equation}
\label{eq:Vb1}
V_b \approx 0.25\,\frac{e}{\,C\,} 
\end{equation}
for $E_J/E_C\ll 1$, and 
\begin{equation}
V_b \propto  
\frac{e}{\,C\,}
\left(\frac{E_J}{E_C}\right)^{\!3/4}\exp\left[-\left(8\,\frac{E_J}{E_C}
\right)^{\!1/2}\,\right]
\end{equation}
for $E_J/E_C\gg 1$, where $C$ is the capacitance of 
the single junction.

The relationship between a superconducting SET 
and a single Josephson junction in the low-impedance environment 
has been discussed,\cite{Joy95}  
and by examining the Hamiltonian, 
it has been shown that a superconducting SET 
can be viewed as a single 
Josephson junction with gate-tunable $E_J$.   
I have found a similar relationship for the 
high-impedance case, which explains the variation 
of $V_b$ in Fig.~\ref{fig:Vb}.   
In the high-impedance environment, not only the island charge 
$Q_1-Q_2$ [$=q$ in Eq.~(\ref{eq:q})]  
but also the total charge $(C_2Q_1+C_1Q_2)/(C_1+C_2)$ 
seen from the outside
contributes to the charging energy,\cite{Ing92} 
where $Q_i$ is the charge on the $i$th junction. 
This is why in Fig.~\ref{fig:diamond}, 
the Coulomb gap is larger in the high-impedance 
environment.  Let us compare Fig.~\ref{fig:diamond}(b) 
with the Coulomb gap for a single junction in the 
high-impedance environment,\cite{Ing92} 
which is $\pm e/2C$.  
In this environment, 
a SET can be viewed as a single junction 
with gate-tunable $C$ or $E_C$, 
where the minimum of $C$ is $\sim0.25(C_1+C_2)$ 
and the maximum is $\sim0.5(C_1+C_2)$.  
I do not know the exact expression for the effective 
$E_J$ of the SET in the high-impedance environment, 
however, the magnitude should be similar to the 
low-impedance case,\cite{Joy95} $\leq  h\Delta_0/8e^2R_n$.  
From Table~\ref{tab:sample} and $C_1+C_2=1$~fF, which was 
obtained from the curve fit  
in Fig.~\ref{fig:IVnormal}, the effective 
$E_J/E_C$ of the SET in Fig.~\ref{fig:Vb} is always 
smaller than 0.04 ($\ll$1).  
Thus, from Eq.~(\ref{eq:Vb1}), our model predicts 
that $V_b$ oscillates between $\sim$0.08 and 
$\sim$0.16~mV in Fig.~\ref{fig:Vb}.  
The order of magnitude is correct and the gate dependence 
is explained.  
Note that the notion of gate-tunable $C$ is a key 
because at $E_J/E_C\ll1$, $V_b$ depends only on $C$.    

\section{Conclusion}
The transport properties of 
single-electron transistors 
(SETs) have been studied in various electromagnetic environments.  
In half of the samples,   
SQUID-array leads were employed in order to realize 
a magnetic-field-tunable environment 
in the superconducting state and a high-impedance 
environment in the normal state.    
I have demonstrated that the effective charging 
energy of SETs in the normal state becomes larger 
in the high-impedance environment 
than in the low-impedance environment.  
In the superconducting state, 
the current modulation by the gate-induced charge 
changed from $e$ periodic to $2e$ periodic 
in SETs with SQUID-array leads, as 
the zero-bias resistance $R_0'$ of the leads 
was increased to be much larger than the quantum 
resistance $R_K\equiv h/e^2\approx26$~k$\Omega$.  
This change in the periodicity suggests that 
SQUID arrays with sufficiently large $R_0'$ 
suppress the contribution 
of quasiparticles to the charge transport.  
When $R_0'\gg R_K$, a sharp Coulomb blockade 
appears in the current--voltage 
characteristics, and the blockade 
voltage $V_b$ varies depending on the gate-induced 
charge.  The variation of $V_b$ is explained 
within a model that treats a superconducting SET 
as a single Josephson junction with gate-tunable 
junction capacitance in the high-impedance environment.

\begin{acknowledgments}
I am grateful to Y. Nakamura 
for fruitful discussion and to M. J. Dalberth 
for proofreading the manuscript.  
This work was supported 
by Special Postdoctoral Researchers Program 
and President's Special Research Grant of RIKEN,   
and MEXT.KAKENHI(15740190).  
Sample fabrication and measurement 
were done at Semiconductors Laboratory 
(closed in March, 2003), RIKEN.  
\end{acknowledgments}

\end{document}